**Chapter 13**

**The application of collagen in advanced wound dressings**

*Giuseppe Tronci*

*Clothworkers' Centre for Textile Materials Innovation for Healthcare, School of Design & Biomaterials and Tissue Engineering Research Group, School of Dentistry, University of Leeds, UK*

**Abstract**

Chronic wounds fail to proceed through an orderly and timely self-healing process, resulting in cutaneous damage with full thickness in depth and leading to a major healthcare and economic burden worldwide. In the UK alone, 200,000 patients suffer from a chronic wound, whilst the global advanced wound care market is expected to reach nearly $11 million in 2022. Despite extensive research efforts so far, clinically-approved chronic wound therapies are still time-consuming, economically unaffordable and present restricted customisation. In this chapter, the role of collagen in the extracellular matrix of biological tissues and wound healing will be discussed, together with its use as building block for the manufacture of advanced wound dressings. Commercially-available collagen dressings and respective clinical performance will be presented, followed by an overview on the latest research advances in the context of multifunctional collagen systems for advanced wound care.

**13.1 Introduction**

Chronic wounds, also referred to as hard-to-heal ulcers, fail to proceed through an orderly and timely self-healing process, resulting in cutaneous damage with full thickness in depth. Chronic wounds typically take longer than 3 months to heal, and can originate from prolonged application of pressure to the skin (pressure ulcers), diabetes-related reduced nerve function and poor blood circulation (diabetic ulcers), improper functioning of venous valves (venous ulcers), and arterial narrowing at the lower extremities (arterial insufficiency ulcers). Clinical complications arising from this pathology include infection, gangrene, haemorrhage and lower-extremity amputations, potentially resulting in permanent disabilities and pain for patients.

Chronic wounds are a major healthcare and economic burden worldwide. In the UK alone, 200,000 patients suffer from a chronic wound, whilst more than 6 million people are affected in the United States of America [1,2]. Due to the increasing rates of diabetes and obesity as well as an ageing population, it is estimated that 1-2% of the general population will develop a chronic wound, and up to 25% of the patients with diabetes will be affected by an ulcer in their lifetime [3]. Consequently, the global advanced wound care market, including advanced wound dressings, negative pressure wound therapy (NPWT), wound care biologics and other products, is expected to reach nearly $11 million in 2022, growing at a CAGR of 5% from 2016 to 2022 [4]. Due to their easy applicability and availability as well as clinical competence, advanced wound dressings accounted for the largest market share of the advanced wound care technology market in 2015. There is therefore growing attention towards the design of advanced dressing devices that can accelerate healing in chronic wounds by recapitulating aspects of the wound healing microenvironment, and that can be customised according to stratified wound populations, to improve patients' quality of life, to reduce healthcare costs and to create patient-friendly solutions.

**13.1.1 Clinical need of advanced wound dressings**

Wound healing is a dynamic process, which proceeds via overlapping phases of inflammation, epidermal restoration, wound contraction and remodelling (Figure 13.1). This process relies on the dynamic interaction of cells, soluble factors and the extracellular matrix (ECM), so that inflammation can rapidly be resolved to allow for the ingrowth of fibroblasts and keratinocytes [5]. Activation of platelets and secretion of inflammatory cytokines, migration of macrophages, fibroblasts and keratinocytes as well as expression of matrix metalloproteinases (MMPs) and growth factors are key to promote wound contraction and closure, ultimately leading to mature ECM and the formation of functional neo-tissue.

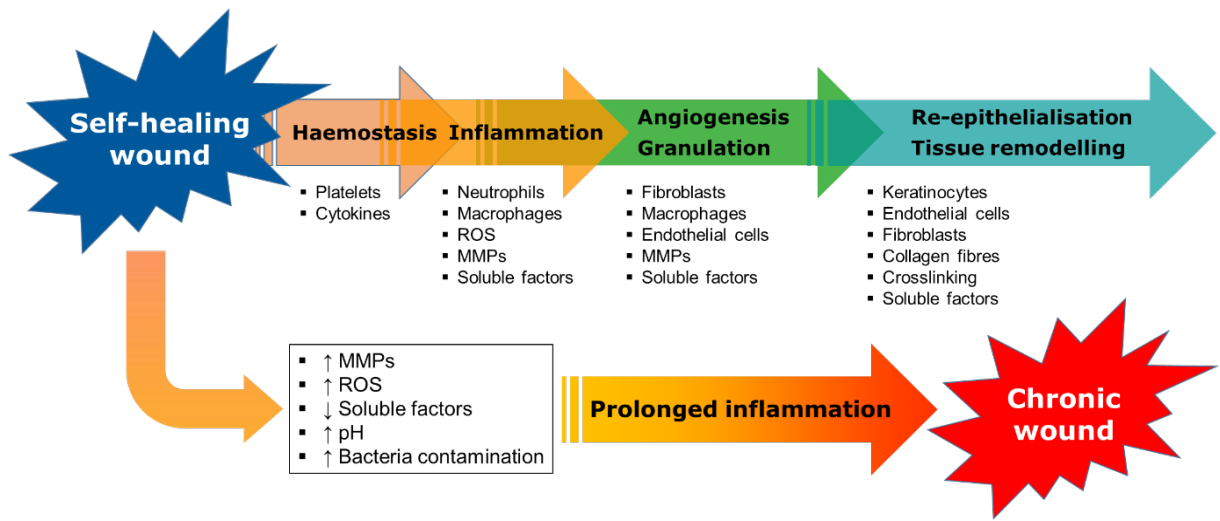

**Figure 13.1.** Typical healing and impaired healing phases in normal and hard-to-heal wounds, respectively.

Especially the release of MMPs in a balanced and coordinated fashion is crucial to enable phagocytosis, angiogenesis, cell migration during epidermal restoration, and tissue remodelling. Such cascade of timely events is impaired in chronic wounds and results in a persistent inflammation state, which is associated with upregulated MMPs and reactive oxidative species (ROS), impaired growth factor expression, and increased risks of bacterial contamination [6]. MMP activity has been reported to be up to 30-fold higher in chronic compared to acute wound fluids, suggesting that new extracellular matrix (ECM) is continuously broken down due to the imbalanced ratio between MMPs and tissue inhibitors of MMPs (TIMPs). In light of the orchestrated nature of the wound microenvironment, observed MMP upregulation negatively affects fibroblast response and differentiation and cause growth factor denaturation at the wound site, so that the subsequent healing steps are halted.

The use of topical wound dressings is a recognised way to wound healing (Figure 13.2). Initially intended to keep the wound dry to minimise wound infection, conventional dressings were subsequently designed to absorb and retain wound exudate in order to enhance healing rates. Highly-hydrated fibrous assemblies have therefore been realised [7], yet a narrow trade-off between dressing exudate absorbency and hydrated mechanical properties is commonly observed *in situ*. Consequently, self-adhesive oxygen-permeable film dressings have been developed to allow for moisture evaporation from intact periwound skin.

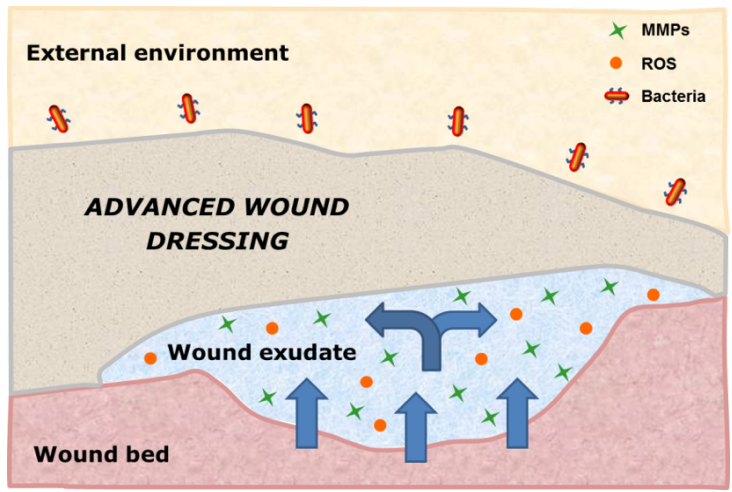

**Figure 13.2.** Design concept of an advanced wound dressing regulating wound exudate levels at the macroscopic scale, as well as pH and overexpressed MMPs and ROS at the biochemical level.

Ultimately, with the advances in biomaterials science, textile manufacture and skin biology, a technology design shift has been pursued from dressings intended as purely skin-protecting bandages to advanced wound dressings, aiming to regulate wound microenvironment with regards to e.g. pH, proteolytic activity and ROS concentrations. An array of approaches of varying efficacy has therefore been pursued, including MMP-cleavable sacrificial substrates [8], metal-chelating chemistries [9], cell [10] and soluble factor [11] delivery strategies, keratinocytes migration-inducing peptides [12], bio-responsive systems [13] and growth factor-delivering vehicles [14]. Despite extensive research efforts so far, clinically-approved therapies for chronic wound management are still time-consuming, economically unaffordable and present restricted customisation.

In this chapter, the role of collagen in the ECM of biological tissues and wound healing will be discussed, together with its use as building block for the manufacture of advanced wound dressings. Commercially-available collagen dressings and respective clinical performance will be presented, followed by an overview on the latest research advances in the context of multifunctional collagen systems for advanced wound care.

**13.2 Collagen as building block of advanced wound dressings**

Collagen is the most abundant protein in the human body. As a structural protein produced by fibroblasts, it plays a major role in all phases of the wound healing cascade, stimulating cellular activity and contributing to new tissue development [15]. The presence of collagen in wound dressings is therefore highly advantageous to promote healing in hard-to-heal wounds, by encouraging the migration of macrophages and fibroblasts to the wound site, leading to the deposition of new collagen matrix. Other than its chemotactic effect on wound specific cells, collagen is also highly hydrophilic, so that respective collagen dressings promote the uptake of growth factors- and MMP-rich wound exudate [16]. Collagen-driven uptake of wound exudate is key not only to keep the wound hydrated, but also aiming to bind and protect wound exudate-carried growth factors, as well as to inactivate upregulated, tissue detrimental chronic wound MMPs, so that neo-tissue is not constantly enzymatically degraded. With the increased elucidation on the role that biochemical factors play in chronic wounds and in light of recent advances in biomaterials science, a great deal of attention has therefore been devoted to the development of advanced collagen-based dressing formulations aiming (i) to provide dressings with superior enzymatic stability at the macroscopic level, (ii) to regulate the chronic wound microenvironment at the biochemical level, and (iii) to successfully enable their customisation to fulfil the requirements of a wide range of chronic wound populations.

**13.2.1 Protein composition and organisation *in vivo***

Collagen accounts for about one-third of the proteins in humans and two-thirds of the skin dry weights. As the most abundant protein in mammals, collagen has provided mankind with widespread applicability. *In vivo*, it plays a dominant role in maintaining the biological and structural stability of various tissues and organs. *Ex vivo*, collagen has been widely employed in the development of leathers and glues, food, cosmetic and pharmaceutical formulations, as well as in medical devices [17]. So far, 28 genetically-distinct types of collagen have been identified, with all of them displaying a triple helical structure at the molecular level (Figure 13.3).

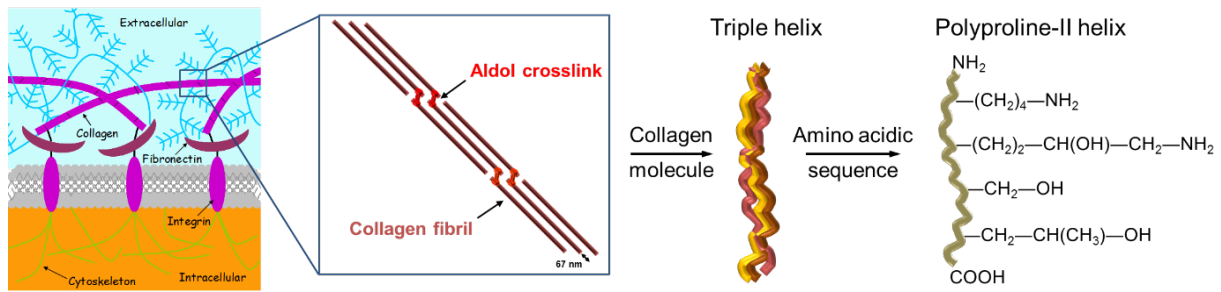

**Figure 13.3.** Stabilised collagen fibril assemblies present in skin's ECM.

Of these, type I (found in skin, tendon, and bone), II (found in cartilage), and III (found in skin and vasculature) are mostly employed for biomedical applications, and are characterised by collagen triple helices assembled into fibrils at the nanoscale, which are responsible for tissue architecture and integrity. The remarkable industrial versatility of collagen can therefore be largely attributed to its hierarchical organisation.

The collagen molecule is based on three left-handed polyproline II-type (PPII) helices, which are staggered from one another by one amino acid residue and are twisted together to form a right-handed triple helix (300 nm in length, 1.5 nm in diameter). Triple helices (THs) can be either homo- or heterotrimers, depending on the tissue and collagen type. With regards to collagen type I, which is widely used in wound care and regenerative medicine, the triple helix consists of a heterotrimer of two α1(I) chains and one α2(I) chain [18]. At the molecular level, each TH-forming PPII helix contains ca. 1000 amino acid residues, and is characterised by the repeating unit *Glycine-X-Y,* whereby *X* and *Y* are predominantly proline and hydroxyproline, respectively. The high content of both stiff (hydroxyl-)proline and small glycine residues explains the folding of each polypeptide into a PPII helix and the consequent arrangement of three PPII helices in the right-handed triple helix. Depending on the location in the human body and the specific biological tissue, THs can assemble into fibrils, fibres and fascicles. Besides secondary interactions, collagen assemblies, i.e. THs, fibrils, fibres and fascicles, are stabilised via covalent crosslinks, which are formed between (hydroxy-)lysine residues under the influence of lysyl oxidase via either aldol condensation or Shiff base-mediated mechanism [19]. These naturally-occurring intra- and intermolecular crosslinks are formed in the non-helical, telopeptide regions of the collagen molecule; they are responsible for the proteolytic resistance of collagen *in vivo* and contribute to the mechanical properties and biological function of tissues.

Above-mentioned chemical composition and structural organisation explains the importance of collagen in wound healing and its widespread use for the manufacture of advanced wound dressings. The presence of the integrin recognition amino acidic sequences RGD and GFOGER along PPII helices enables collagen to control many cellular functions of fibroblasts and keratinocytes, including cell shape, differentiation and migration. Type I collagen has been reported to stimulate angiogenesis *in vitro* and *in vivo* via binding of endothelial cell (EC) surface α1β1 and α2β1 integrin receptors by the $GFPGER_{502-507}$ sequence of the collagen triple helix [20]. The amino acidic sequences of collagen also serve as binding sites for a number of chronic wound-upregulated inflammatory cytokines and MMPs, making collagen biodegradable. Besides promoting the migration of skin cells towards the wound, the application of collagen dressings can also divert tissue detrimental action of aforementioned enzymes and soluble factors from the chronic wound to the dressing, offering an inherent mechanism of wound microenvironment regulation towards healing.

**13.2.2 Extraction of collagen *ex vivo***

Aiming to use collagen as the building block of advanced wound dressings, collagen is extracted from biological tissues, e.g. tendons, in acidic environments or via enzyme-catalysed extraction. Collagen fibers *in vivo* are stable enough to withstand the disruptive influence of thermal agitation but capable of the assembly and disassembly of the component molecules. However, irreversible disassembly of collagen can be induced *ex vivo* by several agents, such as heat, pH, and enzymatic action. In these situations, the weak bonds (hydrogen bonds, dipole-dipole bonds, ionic bonds, and van der Waals interactions) are initially broken, followed by the chemical cleavage of covalent crosslinks. Collagen molecules in the form of triple helices become therefore soluble and diffuse away from the tissue to the extracting medium. Consequently, the unique hierarchical organisation of collagen found *in*

*vivo* is lost *ex vivo*, resulting in a water-soluble product with limited solubility in organic solvents. Extracted collagen triple helices can be reconstituted *in vitro* (pH 7.4, 37 °C) into *in vivo*-like fibrillary structures, resulting in viscoelastic gels at the macroscopic level; yet poor mechanical stability and uncontrollable volumetric swelling are usually observed in aqueous environment.

Whilst the extraction of collagen is typically carried out in mild conditions, so that collagen triple helices can be preserved and collected, excessive heating and extremely acidic solution pH can lead to denatured collagen, whereby triple helices unwind into single random coils, and cleavage of pristine collagen polypeptides into smaller chains occurs. Such denatured form of collagen is called gelatin, which is a heterogeneous mixture of water-soluble polypeptide chains of varied molecular weights. Gelatin usually binds more water than collagen in light of its randomly-coiled conformation, whereby an increase number of functional groups are exposed to water, leading to new hydrogen bonds. Based on their chemical similarities and compatibility within the body, collagen and gelatin are both widely used as building blocks for the design of multifunctional biomaterials and the manufacture of advanced wound dressings. Gelatin is a much cheaper raw material than collagen, and that is why it is often employed in place of collagen. Due to the presence of triple helices and higher molecular weight of PPII helices, collagen shows restricted solubility compared to gelatin, although collagen-based materials typically exhibit increased mechanical competence in hydrated conditions. Consequently, flexible chemical and manufacturing processes should be developed to enable collagen applicability in clinical settings towards the development of *in vivo*-like covalently-crosslinked collagen networks, which are water-insoluble, elastic and that can be configured in bespoke macroscopic formats, such as nonwoven fabrics, coatings as well as pads.

### 13.2.3 Collagen sources and antigenicity

Type I collagen has been successfully extracted from either bovine, equine, or porcine tissues and employed in commercial advanced wound dressings, e.g. Promogran®, Biopad® and Biostep®. However, the use of animal-derived collagen is associated with religious constraints, potential allergies and concerns of transmissible diseases, especially bovine spongiform encephalopathy ("mad cow disease"). Synthetic research approaches leading to either collagen-like macromolecules or collagen triple helix-mimicking peptides have been proposed as chemically viable 'artificial' collagen [21, 22]. Likewise, genetic engineering strategies have been successfully pursued to achieve type I recombinant human collagen as pathogen-free, economically affordable and quality-controlled raw material [23]. On the other hand, alternative collagen sources, such as fish [24,25] and chicken [26] skin, have been explored to address above-mentioned issues in translation and commercialisation settings, and to comply with current regulatory framework. Tissue source is known to affect the extraction yield and the chemical composition and clinical performance of resulting collagen [27], whereby differences in amino acid composition (e.g. lower imino acid and lysine content in fish compared to bovine collagen) have been found to impact on thermal, structural and mechanical properties, covalent crosslinking, and, above all, antigenicity [28]. With regards to the latter point, the use of a human collagen would minimise the probability of interspecies variability and immune rejection, whilst the concept of material purity should also be carefully considered. Therefore, reliable extraction and manufacturing processes should be developed, to minimise the contamination of collagen products with unwanted residues related to non-collagenous proteins, cells, crosslinking compounds, or microbial components, i.e. endotoxins [29].

The selection of collagen sources with minimal antigenicity is prerequisite to enable clinical use of collagen dressings in humans and to minimise the potential to evoke immune and adverse reactions. Macromolecular features present in the collagen backbone not common to the host species are more likely to interact with antibodies and to encourage an immune response than shared features, thereby acting as antigenic determinants. Interspecies amino acidic variation is therefore inherently linked to the issue of collagen antigenicity. Hence, the immunological closeness to humans supports the widespread use of bovine collagen for the development of commercial wound dressings with minimal antigenicity. Antigenic determinants of collagen can be found in the (triple) helical regions, with variations in the amino acid sequences not exceeding more than a few percent between mammalian species [30]. A far greater degree of variability is found in the non-helical terminal regions of collagen, i.e. telopeptides, with up to half of the amino acid residues in these regions exhibiting interspecies variation. Telopeptide-free collagen, also called atelocollagen, can be obtained via pepsin-induced extraction [31], whose yield is reported to be higher compared to acidic extraction [32]. The introduction of pepsin in the extraction medium allows for the selective cleavage of peptide bonds located in the terminal non-helical regions, potentially reducing collagen antigenicity. The telopeptide ends of the collagen molecule are dissected, whilst the triple helices remain preserved. The removal of telopeptides may on the other hand result in the inability of reconstituted product to display characteristic collagen fibril patterns, due to the role amino and carboxyl telopeptides play in crosslinking and fibril formation [28,30].

## 13.3 Commercial collagen dressings

Collagen has been widely investigated as building block of commercial advanced wound dressings, whereby varied protein and chemical configurations have been proposed to achieve superior wound healing dressing function. The application of hydrophilic dressings with appropriate exudate management capability is a recognised route to chronic wound healing [33]. Transparent collagen dressings have been developed in the form of membranes, pads and gels, allowing for wound monitoring and uptake of growth factor-rich wound exudate, whilst providing a barrier to exogenous bacteria. With the increasing understanding of the chronic wound microenvironment, multiphase formulations have also been pursued aiming to integrate collagen-based dressing devices with multiple functions, such as antibacterial activity, drug release as well as wound cleansing and MMP management capabilities (Table 13.1).

**Table 13.1.** Examples of patented, commercially-available collagen-based advanced wound dressings.

| Product ID | Composition | Collagen conformation | Collagen content [wt.%] | Format | Ref. |
|---|---|---|---|---|---|
| Biopad | 100% type I native equine collagen | Triple helix | 100 | Pad | [34] |
| Puracol | 100% bovine collagen, Manuka honey | Triple helix | 88 | Pad | [38] |
| Stimulen | Bovine collagen, glycerin | Hydrolysed | 52 | Gel | [48] |
| Promogran | Bovine collagen, oxidized regenerated cellulose | Hydrolysed | 55 | Pad | [51] |
| ColActive | Gelatin, sodium alginate, carboxymethylcellulose, EDTA, plasticisers | Denatured | 50-90 | Mesh | [53] |

**13.3.1 Non-hydrolysed collagen formulations**

Nonporous dressing pads made of 100% type I equine non-hydrolysed collagen have been developed and commercialised (Biopad®) for the therapeutic treatment of burns and wounds [34]. Here, the hydrophilic behaviour of collagen was exploited to promote absorption (up to 15 times of the material dry weight) of wound exudate and aqueous biological media in the dressing. The dressing product consists of a non-fibrous pad of chemically unmodified collagen, which is prepared by sequential collagen acidic solubilisation, filtering and drying. The fact that the material is nonporous and chemically-unmodified may be beneficial (i) to control and slow down the rapid dressing-induced exudate uptake (due to the presence of a compact, pore-free structure), which may otherwise lead to detrimental and excessive wound evaporation; (ii) and to preserve the native collagen biological function and triple helix structure in resulting product. Especially the retention of triple helices is key to allow for collagen binding with wound exudate growth factors and cytokines and to ensure mechanical competence in the hydrated state [35]. The resulting collagen dressing pad is amenable to surgical cutting to fit the size of the wound. Prior to application *in situ*, the dressing should be partially hydrated to gain elasticity and to conform to the wound, likely due to the glassy-like behaviour of collagen in the dry state and the absence of any plasticising phase in the dressing [36].

Following hydration with wound exudate *in situ*, the collagen dressing is claimed to be transparent, allowing for continuous wound monitoring, so that frequent dressing removal and replacement can in principle be avoided (although a secondary dressing is normally applied to maintain the dressing pad at the wound site). Besides the inherent advantages associated with its 100% non-hydrolysed triple helix-preserved collagen composition, this dressing was also reported to exhibit optimal structural compromise with regards to the extension of the collagen areas and the thickness of the collagen strands. Although no information on the MMP regulation capability was disclosed, these dressings proved to retain the same overall structure during exposure to collagenase [37]. Aiming at a dressing device with additional biochemical and wound cleansing capabilities, a non-hydrolysed collagen-based formulation (Puracol®) was proposed containing Manuka honey (MH) [38]. The use of MH in the dressing is rationalised by the fact that wound cleansing of necrotic tissue is typically required prior to wound dressing treatment; this is usually carried out via surgery or sharp debridement, which may cause patient pain and require additional nursing time. Here, bovine collagen was employed to induce MMP regulation via collagen binding and cleavage with MMPs, and was still applied in its non-hydrolysed, triple helix, and non-crossinked state. Other than controlling MMP overexpression in chronic wounds, the use of collagen was also expected to increase the viscosity of MH aiming to achieve localised delivery to the wound site, minimising MH spillage out of it. The use of MH was first intended to promote debridement of necrotic tissue following dressing application *in situ*, although MH activity is expected to positively impact on wound healing over different levels: (i) its acidity induces temporal denaturation of MMPs, which are responsible for wound chronicity; (ii) MH osmolarity draws wound exudate out of the wound bed, resulting in an outflow of fluid, which helps dissolving necrotic tissue and cleansing the wound [39]; (iii) MH has broad-spectrum antibacterial activity even when in contact with high amounts of wound exudate, largely attributed to the presence of methylglyoxal [40,41].The proposed formulation combining the advantages associated with collagen and MH can be employed as wound contacting layer within a multilayer dressing comprising a (non)woven fabric, an absorbent layer made of e.g. polyurethane foam or cellulose fibres, and a fluid impervious, adhesive cover layer designed to contain the wound exudate absorbed by the dressing. The wound cleansing and healing performance of the collagen layer was successfully investigated in patients with hard-to-heal wounds. Here, decrease in depth, increase in granulation tissue and decrease in the overall wound size provided evidence of the wound conversion from a chronic to a proliferative state within a 3-week time window [42], These findings demonstrated the beneficial function of the native collagen dressing towards the preparation of an optimal wound bed, ultimately leading to accelerated wound re-epithelialisation. On the other hand, the dressing potential to manage MMP upregulation *in situ* was not specifically addressed.

Non-hydrolysed type I equine collagen has also been recently combined with hyaluronic acid (HA) [43] aiming to realise a dressing pad capable to promote cell proliferation, migration, differentiation, and angiogenesis, whilst also inducing hydration of the wound [44]. The use of HA is rationalised since HA is a glycosaminoglycan found in the ECM that plays a central role in controlling water content and mechanical function of connective tissues, whilst also regulating several processes related to cell physiology and biology via interaction with specific cell receptors [45].

Aiming to preserve both collagen proteolytic activity and HA biofunctionality and keep the dressing pad manufacture simple, the formulation is prepared with no covalent crosslinks or chemical functionalisation at the molecular level. Solutions containing specific ratios of HA and collagen are freeze-dried so that the final dressing pad is achieved ready for sterilisation. The application of freeze-drying to the polymer solution enables the formation of pores in resulting dressing internal architecture [46], thereby enhancing the exudate absorption capacity of the dressing (Figure 13.4). Despite the absence of any covalent crosslink between the two phases, carboxylic acid groups of HA are expected to electrostatically interact with the amino groups of collagen, resulting in the formation of a physical network at the molecular scale. In light of the addition of HA, resulting dressing is mechanically strong yet flexible, in contrast to the case of collagen only pad [34], and can promptly conform to topical wounds.

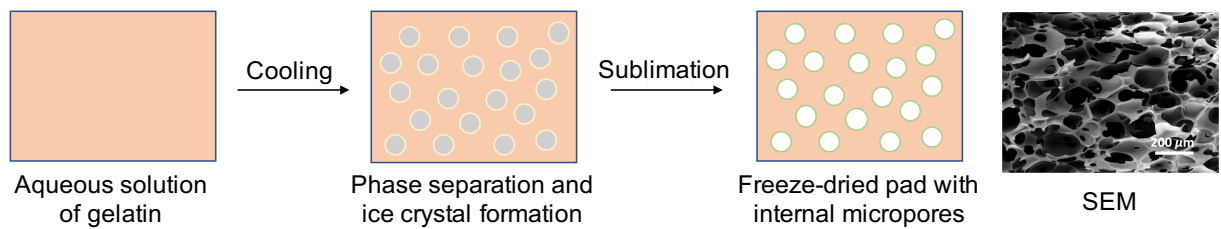

**Figure 13.4.** Manufacture and scanning electron microscopy (SEM) of a dressing pad obtained via freeze-drying of a gelatin aqueous solution.

Other than wound dressing pads, the HA-collagen formulation can also be delivered in the hydrogel state for the treatment of cavity wounds. Culture of 3T3 fibroblasts with both collagen-HA and collagen control formulations confirmed that the use of HA promoted increased adherence of seeded cells to the material with respect to the collagen control, within a 5-day time window. Also in this case, the performance of the HA-collagen dressing prototype in contact with MMPs was not addressed.

### 13.3.2 Hydrolysed collagen-based formulations

Extraction of collagen with preserved triple helices requires defined and controlled experimental conditions. Once extracted *ex vivo*, collagen in its triple helix state is known to present limited solubility and time-consuming solubilisation in aqueous conditions, which limit the creation of scalable customised material format. To overcome above-mentioned constraints and identify cost-effective biochemically-comparable alternatives, extensive research and development has been carried out with derived, either hydrolysed or denatured, forms of collagen, which have been successfully integrated in several commercially-available advanced wound dressings. In contrast to native collagen, hydrolysed or denatured (i.e. gelatin) derivatives consist of a wide mixture of predominantly linear polypeptides. These random coils can however refold into collagen-like triple helices, depending on the environmental conditions and polypeptide molecular weight, so that uncontrollable structure-property relationships may be observed [47]. A modified collagen gel (MCG) formulation (StimulenTM) comprising non-crosslinked, hydrolysed bovine collagen mixture of long and short polypeptides dispersed in a matrix of water and glycerine has been disclosed for the treatment of ischemic wounds [48]. Preclinical investigations in an excisional wound swine model showed that MCG-treated wounds displayed longer rete ridge structures compared with untreated wounds, suggesting improved biomechanical properties of the healing wound tissue. The hydrolysed collagen formulation also proved to improve recruitment of neutrophils and release of inflammation-related cytokines into the wound site [49]. Despite lacking its native triple helix structure, the application of hydrolysed collagen *in situ* was still effective in promoting an initial boost in macrophage concentration and inflammatory response, followed by rapid return of macrophage count to control values and inflammatory resolution, so that healing could take place [50]. Consequent to these results, hydrolysed collagen was blended with oxidised regenerated cellulose (ORC) to create a dressing pad (Promogran®) combining superior water absorption properties of cellulose with wound-regulating biochemical functionalities of collagen [51]. The dressing preparation involved the solubilisation of bovine hydrolysed collagen (BHC) and ORC in an aqueous solution, followed by one-pot freeze-drying and dehydrothermal crosslinking process. The final material therefore consists of a mechanically strong pad, which can be surgically cut to fit the wound size, yet displaying restricted elasticity, likely consequent to the presence of covalently-crosslinked, low molecular weight collagen polypeptides. In contrast to previously-discussed products, results on the BHC/ORC dressing performance in diabetic foot ulcers wound fluid confirmed the material ability to successfully bind and rapidly inactivate chronic wound proteases [52], although no information on the potential dressing mass loss consequent to proteolytic cleavage was provided. Overall, these results underline that the BHC/ORC formulation could remove excess proteases from the wound bed, so that the enzyme-induced tissue degradation could be reduced and tissue synthesis promoted. Despite hydrolysed rather than triple helix preserved collagen was employed in this dressing, these results demonstrate that the collagen polypeptide can still act as a competitive enzymatic substrate with respect to the neo-tissue. Other than the collagen-based mechanism, the addition of ORC in the dressing was demonstrated to provide an additional means for MMP regulation via electrostatic complexation of ORC negatively charged functional groups with positively charged metal ions found in physiological conditions and essential for MMP activity. Hydrolysed collagen was also demonstrated to bind with and protect platelet-derived growth factor (PDGF) from proteolytic degradation, suggesting the potential applicability of the BHC/ORC dressing as controlled delivery system. Moreover, the dressing was also demonstrated to serve as free radical scavenger, which is key to control the excessive upregulation of reactive oxygen species responsible for the prolonged inflammatory state in chronic wounds. The multiple biochemical functions exhibited by the BHC/ORC dressing proved key to induce significantly increased wound closure in diabetic mice in contrast to dressing-free control wounds [8]. These results were found in agreement with histological analysis of wound tissue 14 days post-wounding, whereby enhanced formation and maturation of granulation tissue was observed in BHC/ORC-treated wounds. Other than hydrolysed collagen, porcine gelatin has been employed for the manufacture of flexible mesh dressings, together with a biocompatible plasticiser, such as polyethylene glycol (PEG), carboxymethylcellulose (CMC), sodium alginate and ethylenediaminetetraacetic acid (EDTA) [53]. During manufacture, the dressing-forming aqueous mixture is poured on to a fibrous substrate of e.g. cellulose prior to freeze-drying. The use of the plasticiser in the dressing formulation provides the dressing with enhanced elasticity in the dry state, in contrast to previously mentioned collagen-based dressing pad products, whilst CMC and sodium alginate are employed as hydrophilic component to enhance the absorption capacity of the material *in situ*. Despite gelatin is crosslinked by carbodiimide-induced intramolecular crosslinking, a 20 wt.% mass loss was recorded following 24-hour incubation with collagenase. To further control collagenase activity, EDTA is introduced in the dressing as soluble metal chelator, aiming to induce complexation with the active zinc site of upregulated MMPs, thereby providing an additional mechanism for MMP inactivation. Although the extent of degradation was reduced in the presence of alginate, the structure was observed to collapse following 5-hour exposure to collagenase at body temperature [37], whilst a water uptake of up to 34-time the initial dressing dry weight was observed following 24-hour incubation (0.01 M PBS, pH 7.4, 37 °C). The presence of plasticisers effectively enhanced the elasticity of the material following 1-hour soaking in PBS (0.01 M, pH 7.4), resulting in a measured elongation at break of more than 190%. Ultimately, resulting dressing mesh could be loaded with Sirolimus as exemplary anti-inflammatory drug for the controlled delivery to the surface of tissues, whereby less than 2% drug was released after 3 days *in vitro* (0.01 M PBS, pH 7.4, 37 ºC).

**13.4 Design of multifunctional collagen systems with customised formats**

Despite wound management collagen solutions have been commercialised to respond to the pressing needs of an increasing diabetic population, limitations in biochemical functionalities, manufacturing processes and customisation of dressing properties, functions and format prevent us from developing cost-effective technologies personalised for a wide range of chronic wounds. To address this challenge, widespread research has been pursued to design multifunctional systems with bespoke architecture and integrated bioactive formulations, aiming to correct chronic wound biochemical imbalances and achieve dressing-induced orchestration of the wound healing microenvironment. Research strategies to realise this have focused on three main streams: (i) cell-based therapies, whereby cells are encapsulated in the dressing material to promote tissue repair [54,55]; (ii) drug-loaded systems enabling controlled and staggered release of soluble factors to prevent wound infection and accelerate healing [56,57]; (iii) cell- and soluble factor-free dressing devices whereby the healing functionality is inherently accomplished by dressing physical properties and chemical configurations at the molecular scale [58,59]. The design of inherently multifunctional wound dressings may be preferable to achieve constant clinical performance *in vivo*, irrespective of temporal factors ruling e.g. the release of active compounds or metabolic activity of encapsulated cells, whilst also minimising regulatory framework constraints and time required for regulatory approval and translation to market.

**13.4.1 Synthesis of network architectures to achieve dressing multifunctionality**

Despite collagen is an ideal biomaterial for wound dressing application, the inherent MMP-induced degradation *in vivo* of collagen-based dressings raises concerns in terms of dressing form-stability, non-controllable swellability, and poor hydrated mechanical properties. Incorporation of soluble MMP-chelating agents, i.e. EDTA, has been exploited for the development of commercial advanced wound care products. Yet, loading of the dressing with soluble factors may involve additional considerations with regards to controlled release kinetics, sustained MMP modulation and medicinal product-related translation pathway. Either glutaraldehyde [60,61], genipin [62], carbodiimide chemistry [63], or physical factors [64,65] have been employed to stabilize collagen in biological environments, although issues remain with regards to material cytotoxicity and restricted control of crosslinking reaction and material properties. To overcome these limitations, a great deal of attention has been given towards the development of flexible crosslinking strategies of collagen. By introducing crosslinks between collagen molecules, water-insoluble networks are accomplished that swell in contact with wound exudate, so that defined moist environment is established *in situ*, whilst enhanced dressing biodegradability and hydrated mechanical properties can be expected. Most importantly, additional biofunctionalities can be introduced in resulting materials depending on the characteristics of the crosslinking segment.

Type I bovine collagen has been crosslinked with tannic acid in an effort to introduce the antimicrobial and anti-inflammatory activities of tannic acid within a covalent collagen matrix [66]. Tannic acid (TA) is a plant polyphenol consisting of a glucose moiety core with hydroxyl groups being esterified with five digallic acids. Resulting materials proved to display about 10 wt.% mass loss following 42-hour incubation in a collagenase-containing medium, in contrast to more than 60 wt.% mass loss observed in tannic acid-free collagen controls in the same conditions. Following application to full-thickness wounds in rats, TA-crosslinked samples proved to support enhanced wound closure and nearly-complete re-epithelialisation following 12-day treatment, in comparison with the collagen control, whilst no commercial benchmark was employed in the study. Results obtained *in vitro* and *in vivo* therefore support the hypothesis that TA carboxylic and hydroxyl groups within the collagen matrix can mediate the formation of non-covalent net-points with the functional groups of the collagen molecules, explaining the decreased degradation yield. Digallic acid units are also known to act as metal ion chelators; they can therefore bind with and inhibit collagenases [67], contributing to the accelerated healing observed in wounds treated with TA-based collagen formulations. Building on the knowledge developed with this system, Francesko et al. analysed the enzymatic modulation activity of plant polyphenol-loaded type I bovine collagen-based co-networks prepared via carbodiimide-catalysed crosslinking reaction with either hyaluronic acid or chitosan [68]. Hydrogels displayed an averaged swelling ratio of 1500-2500 wt.% and an averaged Young's modulus of 60-160 kPa, depending on the specific formulation. Here, the addition of hyaluronic acid proved to significantly impact on the water uptake capability of resulting samples, in agreement with the well-known swellability of hyaluronic acid *in vivo*. Despite that, increased material stability was observed in enzymatic media, whereby collagen degradation was measured in the range of 10-30 wt.% following 1-day incubation with collagenases. The release profile of loaded polyphenols was not addressed, yet no toxic response and spread-like cell morphology was observed following 3-day culture with L929 fibroblasts. Following similar line of thinking, either *Macrotyloma uniflorum* or *Triticum aestivum* were loaded as anti-inflammatory and antibacterial plant extract onto fish collagen-fibrin composites [69] and goat tendon collagen aerogels [70], respectively. Collagen-based composites displayed more than 40 wt.% mass loss following 24-hour incubation in enzymatic media, partially explained by the absence of any covalent crosslinks; yet, they induced accelerated healing of full thickness wounds in albino Wistar rats via suppression of cyclooxygenase-2 (COX-2), inducible nitric oxide synthases (iNOS) and MMP-9 expressions [71]. On the other hand, *triticum aestivum* was shown to promote chemical crosslinking of collagen lysines [70], so that *triticum aestivum*-loaded collagen aerogels displayed proangiogenic effect resulting in complete wound closure in female wistar rats following 18 days post-wounding. Besides the biochemical modulation of the wound microenvironment, sponges of denatured collagen and hyaluronic acid were realised and loaded with epidermal growth factor (EGF), aiming to stimulate cell proliferation for the treatments of burns [72]. The absence of covalent crosslinks between the two biopolymers was likely responsible for the quick dissolution of the sponge within 7-day incubation, although the consequent quick EGF release proved beneficial in promoting angiogenesis and re-epithelialisation in a dermal burn model in rats [73] as well as in a full-thickness dorsal skin defect in diabetic mice [74]. Other than loading with soluble bioactives to achieve specific modulation of the wound microenvironment, a family of functionalised (atelo)collagen networks has recently been developed, whereby inherent control of MMP activity is accomplished via the introduction of chemically-coupled photoactive compounds [75,76].

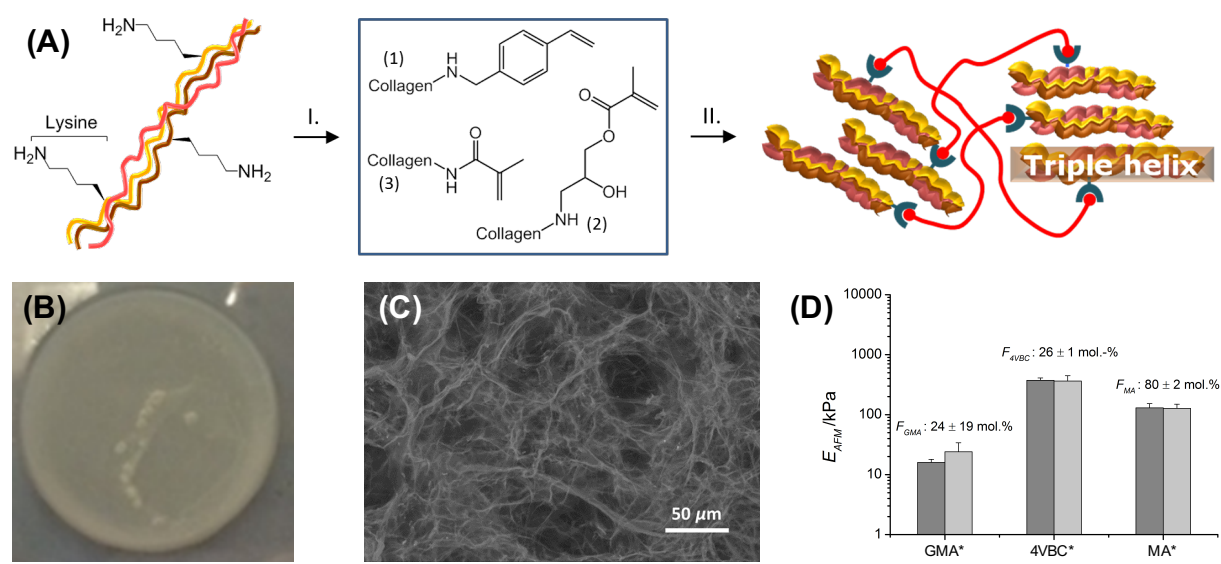

**Figure 13.5.** (A): Synthesis of UV-cured (atelo)collagen networks via lysine functionalisation with photoactive compounds. (B): Resulting (atelo)collagen networks swell in water and present an internal porous architecture (C) by SEM. (D): Atomic force microscopy on collagen hydrogels reveal significantly different elastic modulus ($E_{AFM}$) depending on the specific network architecture.

The covalent functionalisation of (atelo)collagen triple helices with photoactive compounds enables (i) the formation of photo-induced covalent networks with bespoke macroscopic properties and (ii) the complexation of introduced adducts with the active site of MMPs, yet avoiding the use of any soluble factor. This synthetic strategy proved to enable customisation of hydrogel elastic modulus depending on the type of chemically-coupled photoactive adduct (Figure 13.5), whilst an averaged swelling ratio of up to nearly 2000 wt.-% was observed [77]. Preclinical investigations with a Hydrogel of Functionalised ateloCollagen (HyFaCol) indicated increased neodermal response in full thickness wounds created in diabetic mice, compared to wounds treated with a polyurethane commercial control in the same experimental conditions [78].

**13.4.2 Customisation into single fibrous component for wound dressing manufacture**

The clinical performance of multifunctional collagen-based formulations would be greatly accelerated if delivered in material formats relevant to the wound dressing manufacturing industry. Fibrous architectures, e.g. nonwovens, are widely used for the development of healthcare materials, including wound dressings, due to their high porosity, easy manufacture and advantageous fluid adsorption properties [79]. Consequently, the creation of single collagen fibres has been widely pursued aiming to accomplish libraries of dressing building blocks with customised molecular architecture, properties and biofunctionalities, to fulfil the complex requirements of stratified chronic wounds. To deliver on this vision, specific manufacturing routes need to be developed enabling preservation of collagen proteinic architecture in the fibrous state, whilst allowing for clinically-acceptable material purity and manufacturing yield relevant for industrial scale up.

Among the different fibre manufacturing processes available, wet spinning has the potential to convert collagen solutions into single fibres. From a protein preservation standpoint, this fibre spinning process is highly benign since there is no involvement of either harsh organic solvents, electrostatic voltage or high temperature [80]. In these experimental conditions, the risk of spinning-induced collagen triple helix denaturation is minimised, so that wet spun fibres consisting of preserved collagen triple helices can be obtained. Wet spinning was developed by the textile industry in the early 1900s as a means of producing man-made fibres such as viscose rayon. Mechanistically, this fibre spinning mechanism relies on (i) the phase separation of a fibre-forming (bio)polymer solution against a suitable (bio)polymer non-solvent and on (ii) the spinning rate of the phase-separating solution jet. The (bio)polymer solution is extruded through a spinneret into a non-solvent coagulation bath, whereby (bio)polymer solution streams turn into solid filaments, due to the non-solvent-induced (bio)polymer phase separation. For each (bio)polymer system, fibre properties can therefore be manipulated by adjusting solution and coagulating bath characteristics, wet-spinning parameters (e.g. spinneret diameter, solution injection flow rate, drawing ratio) and post-spinning fibre conditioning (e.g. covalent crosslinking, washing, drying).

Silver et al. pioneered the formation of high strength, dehydrothermally-crosslinked, wet spun collagen fibres using concentrated rat tail collagen solutions in diluted hydrochloric acid [81,82]. Wet spinning was carried out in a neutral buffer system at 37 ºC, whereby resulting fibres were washed in isopropyl alcohol and distilled water prior to dehydrothermal crosslinking. Remarkably, hydrated wet spun fibres exhibited a maximum tensile strength and elongation at break of up to 92 MPa and 20%, respectively, although the density of the covalent crosslinks could not be quantified. Using a PEG-containing coagulating bath, Zeugolis et al. successfully wet spun bovine atelocollagen fibre with *in vitro* reconstituted fibrillary organisation and increased tensile strength [83]. Post-spinning incubation in isopropanol induced a decrease in fibre diameter compared to fibres conditioned in aqueous environment, likely related to fibre dehydration in the former case and water-induced fibre swelling in the latter case. Hydrated wet spun fibres displayed an elastic modulus of more than 16 MPa when incubated in PBS, and of nearly 4 MPa when incubated in distilled water, suggesting that PBS incubation induced folding of collagen triple helices into fibrils. In another account, wet spun collagen was incubated with crosslinking agents, resulting in fibres resembling the tensile properties of native tissues. Either ethylene glycol diglycidyl ether and hexamethylene diisocyanate (HDI) proved suitable in enhancing the mechanical properties of wet spun collagen, although concerns were also raised regarding potential material cytotoxicity and side reactions [47]. To minimise the extent of superficial irregularities, e.g. ridges and crevices, following fibre air drying and increase molecular alignment and fibre tensile properties, Caves et al. applied drawing to wet spun fibres. A dual syringe pump system was developed where isolated collagen solution and PEG-based wet spinning buffer were mixed in 1-m fluoropolymer tubing before entering in a 2-m ringing bath of 70% ethanol solution [84]. Resulting filaments displayed retained triple helix organisation, whilst reconstituted collagen fibrils were also observed following 48-hour incubation in PBS at 37 ºC and washing in distilled water, prior to fibre crosslinking with glutaraldehyde. Hydrated fibres displayed a fibre diameter in the range of 20-50 μm and an averaged ultimate tensile strength of nearly 94 MPa. Treatment with glutaraldehyde proved successful in decreasing the extent of fibre degradation following subcutaneous implantation in C57BL/6 mice, although a mild local inflammatory response was observed. Integrated wet spinning-crosslinking processes have also been attempted, whereby fibre-forming collagen was wet spun (with no drawing) in a mineral salt aqueous coagulation bath containing glutaraldehyde [85]. Despite grooved surface morphology was observed, resulting fibres could be successfully assembled in a carded nonwoven architecture. This study therefore supported the use of wet spinning for the formation of fibrous assemblies based on triple helix preserved collagen, in contrast to the use of e.g. melt spinning, whereby fibres of completely denatured collagen are obtained [86]. Other than animal-derived collagen, recombinant human type I collagen derived from a transgenic tobacco plant was recently applied in a novel fibre spinning method, involving drawing and crosslinking [23]. Smooth, highly aligned wet spun fibres with a fibre diameter as small as 8 μm could be successfully accomplished. Also in this case, potentially toxic glutaraldehyde and carbodiimide-induced crosslinking was pursued, whereby the former treatment led to fibres with increased wet-state stress at break (~140 MPa) compared to carbodiimide-crosslinked fibres (~40 MPa). To address the cytotoxicity issues associated with glutaraldehyde or carbodiimide, a photoactive collagen system has been recently developed and employed as fibre building block [87]. Photoactive collagen proved to be compatible with the wet spinning process in either alcohol or water-based coagulating baths. Water-insoluble, mechanically-competent fibres could be prepared via post-spinning UV-curing process, and successfully assembled into a three-dimensional porous fabric (Figure 13.6). Other than wet spinning, microfluidics-based approaches have also been proposed for the fabrication of reconstituted type I collagen fibres with fibre diameters of only 3 μm and an averaged tensile strength of 383 MPa [88].

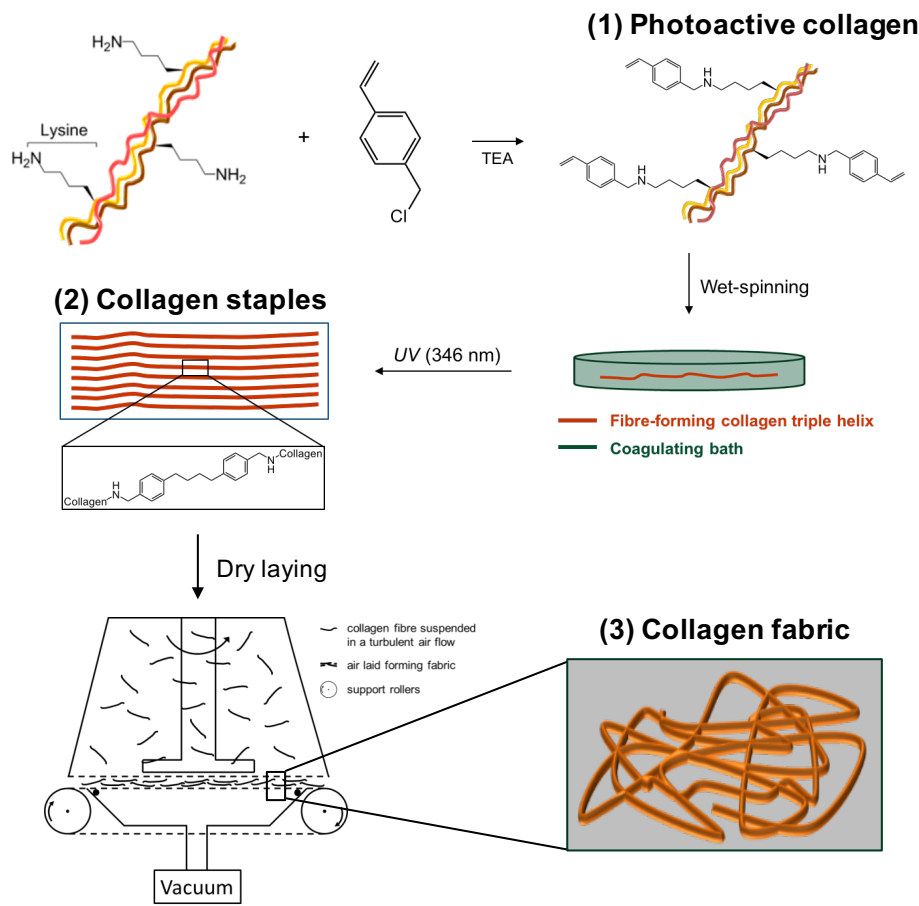

**Figure 13.6.** Manufacture of a collagen fabric with retained triple helices. 4VBC-functionalised collagen (1) is dissolved in acidic solution and wet spun. UV-cured staples (2) are water insoluble and mechanically competent, and can be assembled into a three-dimensional fabric via e.g. dry laying (3).

Despite the remarkable performance of resulting fibres and the absence of crosslinking step, further research may be required into this process to increase its 19 m·h$^{-1}$ production rate towards wet spinning-like fibre production rates of 1000 m·h$^{-1}$ [23] to enable industrial scale up.

### 13.4.3 Direct manufacture of multifunctional collagen-based meshes

Together with the multiscale manufacture of fabric dressings via controlled customisation of multifunctional molecular systems into single fibres and consequent fibre assembly into three dimensional structures, one-step spinning processes have also been extensively investigated for the creation of fibrous meshes from collagen-based solutions. Such research approaches undoubtedly offer a faster manufacturing route to the fibrous prototype, although reduced controlled over fibre characteristics, protein organisation and production yield is expected [36,80,86]. Electrospinning and wet electrospinning of collagen-based materials have been extensively employed for the formation of either mono- or multi-layered nonwoven fabrics, whilst other strategies involving physical or chemical deposition of collagen coatings onto fibrous supports have also been pursued.

Electrospinning is considered as a simple and effective fabrication method to prepare nanofibrous membranes with diameters ranging from 5 to 500 nm, i.e. about 100 times smaller than the fibre diameters observed in wet or melt spun fibres. Electrospinning involves the application of voltage to a syringe containing a polymer solution [89]. Following ejection of the polymer solution and solvent evaporation, a nonwoven mesh is formed on a grounded collector. Despite fibre and web characteristics can be manipulated by adopting appropriate experimental parameters, such as solvent, polymer concentration, and flow rate, common process limitations include the limited fabric thickness and restricted fabric porosity. Furthermore, given that electrospinning requires volatile solvent to allow for prompt fibre formation, organic solvents are usually. In the case of collagen, both applications of electrostatic voltage and organic solvents are well known to induce unwinding of collagen triple helices into randomly-coiled polypeptide chains, so that resulting mesh is effectively composed of gelatin rather than collagen [90,91]. Due to the loss of collagen triple helices, resulting samples are readily soluble in aqueous environment, preventing their use in biological environments. Consequently, the formation of water-stable collagen-based electrospun membranes has attracted widespread research attention and has so far been typically accomplished via blending of collagen solution with other polymers [35], manufacture of multi-layered structures [92,93,94], chemical crosslinking of resulting fibres [95,96,97], or combination thereof. These approaches proved successful towards the creation of ECM-mimetic fabrics integrated with growth factor retention and release functionality [98]. Here, a polysaccharide component allowed for the effective sequestration of endogenous platelet-derived growth factor; whilst, the enzymatic cleavage of gelatin in the electrospun matrix triggered the release of sequestered growth factor, so that accelerated repair of a full-thickness skin wound was observed in C57BL/6 mice.

To overcome the cytotoxicity issues with conventional crosslinking methods, e.g. glutaraldehyde, Dhand et al. have recently proposed the addition of naturally occurring catecholamines in collagen-based electrospinning solutions, so that fibre stabilisation is accomplished via latent oxidative polymerization initiated by exposure to ammonium carbonate [99]. In another recent report, wet electrospinning has been proposed to address the limited thickness, porosity and pore size observed in electrospun fabrics [100]. Wet electrospinning proceeds by spraying electrospun nanofibers into a liquid bath, which is used to separate the fibers. By controlling the area of nanofiber deposition, resulting wet electrospun webs revealed a porosity of about 90 vol.%, in contrast to 60-80 vol.% porosity observed in electrospun samples. Other approaches have also been investigated to avoid inherent electrospinning-associated denaturation of triple helices. Collagen has been deposited as physical coating onto polyester/gelatin electrospun meshes, yet rapid collagen degradation was observed during 24-hour enzymatic incubation. Chemical immobilisation of collagen coating has therefore been proposed by using polypropylene (PP) meshes as inert substrate for biomimetic functionalisation [101]. Plasma modification of PP was applied to enable the incorporation of acrylic moieties, which could then be activated with carbodiimide to covalently couple collagen. Although the enzymatic stability of the coating was not addressed, application of resulting material on to a full thickness wound model in SD rats revealed increased wound closure with respect to wounds treated with PP controls. Similar approach was also pursued in an N-isopropyl acrylamide (NIPAM)-grafted PP nonwoven fabric, whereby covalent glutaraldehyde-mediated covalent linkages were proposed between collagen and NIPAM was proposed [102].

**13.5 Outlook**

Different advanced wound care concepts will continue to be developed to support chronic wound healing, which will rely on new antibiofilm technologies, multifunctional material design, as well as flexible and scalable manufacture. A strong focus will be on integrated collagen-based dressings that will allow remote monitoring of wound conditions and visual indications of chronic state changes. This approach will enable timely and prolonged application of cost-effective dressing devices promptly customised to the targeted chronic wound, ultimately resulting in economically-affordable wound healing times, and minimised dressing changes and risks of infection. Flexible design and manufacturing concepts should therefore be developed, which can enable systematic variations in dressing properties and functions, whilst also allowing for late stage device assembly at the bed side. Especially in out-of-hospital healthcare, collagen devices that are able to adjust to and regulate changes in chronic wound microenvironment, whilst also displaying easy removability, are expected to reduce patient pain, therapeutic time and wound infection risks. These requirements could be fulfilled by systems that are able to perform and display more than one function according to defined biochemical shifts. Advances in collagen-compatible fibre spinning and assembling technologies will be paramount to realise high-value building blocks with integrated sensing units and to enable the conversion into appropriate structures depending on the chronic wound type, state and size. Patient- and clinician-assisted device development will continue to be key to de-risk late-stage clinical failure and to identify appropriate clinical models for first-in-man evaluations, prior to clinical trials.

**Acknowledgments**

The author gratefully acknowledges financial support from the Clothworkers' Centre for Textile Materials Innovation for Healthcare (CCTMIH), the EPSRC-University of Leeds Impact Acceleration Account (IAA), and the EPSRC Centre for Innovative Manufacturing in Medical Devices (MeDe Innovation Fresh Ideas Fund).

# References

[1] A. Sharpe, M. Concannon, Demystifying the complexities of wound healing, Wounds UK 8 (2) (2012) 81-86

[2] K. Järbrink, G. Ni, H. Sönnergren, A. Schmidtchen, C. Pang, R. Bajpai, J. Car, Prevalence and incidence of chronic wounds and related complications: a protocol for a systematic review, Systematic Reviews 5 (152) (2016), 1-6

[3] C.C.L.M. Naves, The Diabetic Foot: A Historical Overview and Gaps in Current Treatment, Advances in Wound Care 5 (5) (2016) 191-197

[4] Global Advanced Wound Care Market - Analysis and Forecast (2016-2022) (Focus on Advanced Wound Care Dressings, NPWT Devices, HBOT Devices, Wound Care Biologics, Ultrasonic Devices, and Electromagetic Devices). Report ID: 4717465. <https://www.reportbuyer.com/product/4717465/>, 2016

[5] J. Ho, C. Walsh, D. Yue, A. Dardik, U. Cheema, Current Advancements and Strategies in Tissue Engineering for Wound Healing: A Comprehensive Review, Advances in Wound Care 6 (6) (2017) 191-209

[6] I. Stefanov, S. Pérez-Rafael, J. Hoyo, J. Cailloux, O.O. Santana Pérez, D. Hinojosa-Caballero, T. Tzanov, Multifunctional Enzymatically Generated Hydrogels for Chronic Wound Application, Biomacromolecules 18 (5) (2017) 1544-1555

[7] M.J. Waring, D. Parsons, Physico-chemical characterisation of carboxymethylated spun cellulose fibres. Biomaterials 22 (9) (2001) 903-912

[8] J. Hart, D. Silcock, S. Gunnigle, B. Cullen, N.D. Light, P.W. Watt. The role of oxidised regenerated cellulose/collagen in wound repair: effects in vitro on fibroblast biology and in vivo in a model of compromised healing. The International Journal of Biochemistry & Cell Biology 34 (12) (2002) 1557-1570

[9] M. Gooyit, Z. Peng, W.R. Wolter, H. Pi, D. Ding, D. Hesek, M. Lee, B. Boggess, M.M. Champion, M.A. Suckow, S. Mobashery, M. Chang, A Chemical Biological Strategy to Facilitate Diabetic Wound Healing, ACS Chemical Biology 9 (1) (2014) 105-110

[10] G. Eke, N. Mangir, N. Hasirci, S. MacNeil, V. Hasirci, Development of a UV crosslinked biodegradable hydrogel containing adipose stem cells to promote vascularization for skin wounds and tissue engineering, Biomaterials 129 (2017) 188-198

[11] H. Hathaway, J. Ajuebor, L. Stephens, A. Coffey, U. Potter, J.M. Sutton, A.T.A. Jenkins, Thermally triggered release of the bacteriophage endolysin CHAPK and the bacteriocin lysostaphin for the control of methicillin resistant Staphylococcus aureus (MRSA), Journal of Controlled Release 245 (2017) 108-115

[12] Y. Xiao, L.A. Reis, N. Feric, E.J. Knee, et al. (2016). Diabetic wound regeneration using peptide-modified hydrogels to target re-epithelialization. Proc. Natl. Acad. Sci. USA 113 (2016) E5792-E5801

[13] B.P. Purcell, D. Lobb, M.B. Charati, S.M. Dorsey, R.J. Wade, K.N. Zellars, H. Doviak, S. Pettaway, C.B. Logdon, J.A. Shuman, P.D. Freels, J.H. Gorman, R.C. Gorman, F.G. Spinale, J.A. Burdick, Injectable and bioresponsive hydrogels for on-demand matrix metalloproteinase inhibition, Nature Materials 13 (2014) 653-661.

[14] J.T. Hardwicke, J. Hart, A. Bell, R. Duncan, R. Moseley. The effect of dextrin–rhEGF on the healing of full-thickness, excisional wounds in the (db/db) diabetic mouse. Journal of Controlled Release 152 (2011) 411-417

[15] T. Zhou, N. Wang, Y. Xue, T. Ding, X. Liua, X. Mo, J. Sun. Electrospun tilapia collagen nanofibers accelerating wound healing via inducing keratinocytes proliferation and differentiation. Colloids and Surfaces B: Biointerfaces 143 (2016) 415-422.

[16] M. Motzkau, J. Tautenhahn, H. Lehnert, R. Lobmann. Expression of matrix-metalloproteases in the fluid of chronic diabetic foot wounds treated with a protease absorbent dressing. Experimental and Clinical Endocrinology & Diabetes 119 (2011) 286-290

[17] S. Chattopadhyay, R.T. Raines. Collagen-based biomaterials for wound healing. Biopolymers 101 (2014) 821-833

[18] E.A.A. Neel, L. Bozec, J.C. Knowles, O. Syed, V. Mudera, R. Day, J.K. Hyun. Collagen -- Emerging collagen based therapies hit the patient. Advanced Drug Delivery Reviews 65 (2013) 429-456

[19] R.C. Siegel. Collagen crosslinking. Journal of Biological Chemistry 251 (1976) 5786-5792

[20] J.D. San Antonio, J.J. Zoeller, K. Habursky, K. Turner, W. Pimtong, M. Burrows, S. Choi, S. Basra, J.S. Bennett, W.F. DeGrado, R.V. Iozzo. A key role for the integrin $\alpha2\beta1$ in experimental and developmental angiogenesis. The American Journal of Pathology 175 (2009) 1338-1347

[21] S.E. Paramonov, V. Gauba, J.D. Hartgerink. Synthesis of Collagen-like Peptide Polymers by Native Chemical Ligation. Macromolecules 38 (2005) 7555-7561

[22] K. Kar, P. Amin, M.A. Bryan, A.V. Persikov, A. Mohs, Y.-H. Wang, B. Brodsky. Self-association of Collagen Triple Helix Peptides into Higher Order Structures. Journal of Biological Chemistry 281 (2006) 33283-33290

[23] A. Yaari, Y. Schilt, C. Tamburu, U. Raviv, O. Shoseyov. Wet Spinning and Drawing of Human Recombinant Collagen. ACS Biomaterials Science and Engineering 2 (2016) 349-360

[24] Q.F. Dang, H Liu, J.Q. Yan, C.S. Liu, Y. Liu, J. Li, J.J. Li. Characterization of collagen from haddock skin and wound healing properties of its hydrolysates. Biomedical Materials 10 (2015) 015022

[25] R.C.G. Coelho, A.L.P. Marques, S.M. Oliveira, G.S. Diogo, R.P. Pirraco, J. Moreira-Silva, J.C. Xavier, R.L. Reis, T.H. Silva, J.F. Mano. Extraction and characterization of collagen from Antarctic and Sub-Antarctic squid and its potential application in hybrid scaffolds for tissue engineering. Materials Science and Engineering C 78 (2017) 787-795

[26] Y.Y. Peng, V. Glattauer, J.A.M. Ramshaw, J.A. Werkmeister. Evaluation of the immunogenicity and cell compatibility of avian collagen for biomedical applications. Journal of Biomedical Materials Research Part A 93A (2010) 1235-1244

[27] R. Parenteau-Bareil, R. Gauvin, S. Cliche, C. Gariépy, L. Germain, F. Berthod. Comparative study of bovine, porcine and avian collagens for the production of a tissue engineered dermis. Acta Biomaterialia 7 (2011) 3757-3765

[28] S. Majumdar, Q. Guo, M. Garza-Madrid, X. Calderon-Colon, D. Duan, P. Carbajal, O. Schein, M. Trexler, J. Elisseeff. Influence of collagen source on fibrillar architecture and properties of vitrified collagen membranes. Journal of Biomedical Materials Research Part B: Applied Biomaterials 104B (2016) 300-307

[29] Y. Nakagawa, T. Murai, C. Hasegawa, M. Hirata, T. Tsuchiya, T. Yagami, Y. Haishima. Endotoxin contamination in wound dressings made of natural biomaterials. Journal of Biomedical Materials Research Part B: Applied Biomaterials 66B (2003) 347-355

[30] A.K. Lynn, I.V. Yannas, W. Bonfield. Antigenicity and Immunogenicity of Collagen. Journal of Biomedical Materials Research Part B: Applied Biomaterials 71B (2004) 343-354

[31] R. Holmes, S. Kirk, G. Tronci, X. Yang, D. Wood. Influence of telopeptides on the structural and physical properties of polymeric and monomeric acid-soluble type I collagen. Materials Science and Engineering C 77 (2017) 823-827

[32] W. Friess. Collagen – biomaterial for drug delivery. European Journal of Pharmaceutics and Biopharmaceutics 45 (1998) 113-136

[33] M. Romanelli, K. Vowden, D. Weir. Exudate Management Made Easy. Wounds International 1 (2010) 1-6

[34] D. Furlan, G. Bonfanti, G. Scappaticci. Non-porous collagen sheet for therapeutic use, and the method and apparatus for preparing it. US Patent. #5785983 (1993)

[35] X. Qiao, S.J. Russell, X. Yang, G. Tronci, D.J. Wood. Compositional and in vitro evaluation of nonwoven type I collagen/poly-dl-lactic acid scaffolds for bone regeneration. Journal of Functional Biomaterials 6 (2015) 667-686

[36] G. Tronci, R.S. Kanuparti, M.T. Arafat, J. Yin, D.J. Wood, S.J. Russell. Wet-spinnability and crosslinked fibre properties of two collagen polypeptides with varied molecular weight. International Journal of Biological Macromolecules 81 (2015) 112-120

[37] J.C. Karr, A.R. Taddei, S. Picchietti, G. Gambellini, A.M. Fausto, F. Giorgi. A Morphological and Biochemical Analysis Comparative Study of the Collagen Products Biopad, Promogram, Puracol, and Colactive. Advances in Skin Wound Care 24 (2011) 208-216.

[38] D. Chakravarthy, A.J. Ford. Wound dressing containing polysaccharide and collagen. US Patent App. US14/031,716 (2015).

[39] P. Molan, T. Rhodes. Honey: A Biologic Wound Dressing. Wounds 27 (2015) 141-151

[40] S.E.L. Bulman, G. Tronci, P. Goswami, C. Carr, S.J. Russell. Antibacterial Properties of Nonwoven Wound Dressings Coated with Manuka Honey or Methylglyoxal. Materials 10 (2017) 954

[41] S.E.L. Bulman, G. Tronci, P. Goswami, S.J. Russell, C. Carr. Investigation into the potential use of poly(vinyl alcohol)/methylglyoxal fibres as antibacterial wound dressing components. Journal of Biomaterials Applications 29 (2015) 1193-1200

[42] N. Wahab, M. Roman, D. Chakravarthy, T. Luttrell. The Use of a Pure Native Collagen Dressing for Wound Bed Preparation Prior to Use of a Living Bi-layered Skin Substitute. Journal of the American College of Clinical Wound Specialists 6 (2015) 2-8

[43] S. Srivastava, S.D. Gorham, D.A. French, A.A. Shivas, J.M. Courtney. In vivo evaluation and comparison of collagen, acetylated collagen and collagen/glycosaminoglycan composite films and sponges as candidate biomaterials. Biomaterials 11 (1990) 155-161

[44] G. Gennari, S. Panfilo, J.F. Scalesciani. Pharmaceutical compositions comprising collagen and sodium hyaluronate. US Patent Application US14/894,266 (2016)

[45] D. Campoccia, P. Doherty, M. Radice, P. Brun, G. Abatangelo, D.F. Williams. Semisynthetic resorbable materials from hyaluronan esterification. Biomaterials 19 (1998) 2101-2127

[46] H.-W. Kang, Y. Tabata, Y. Ikada. Fabrication of porous gelatin scaffolds for tissue engineering. Biomaterials 20 (1999) 1339-1344

[47] G. Tronci, A.T. Neffe, B.F. Pierce, A. Lendlein. An entropy–elastic gelatin-based hydrogel system. Journal of Materials Chemistry 20 (2010) 8875-8884

[48] E.I. Stout, C. Sen. Composition and methods for treating ischemic wounds and inflammatory conditions. International Patent Application WO2016109722 A1 (2016)

[49] H. Elgharably, S. Roy, S. Khanna, M. Abas, P. DasGhatak, A. Das, K. Mohammed, C.K. Sen. A modified collagen gel enhances healing outcome in a preclinical swine model of excisional wounds. Wound Repair and Regeneration 21 (2013) 473-481

[50] S.A. Eming, T. Krieg, J.M. Davidson. Inflammation in Wound Repair: Molecular and Cellular Mechanisms. Journal of Investigative Dermatology 27 (2007) 514-525

[51] P.W. Watt, W. Harvey, D. Wiseman, N. Light, L. Saferstein, J. Cini. Wound dressing materials comprising collagen and oxidized cellulose. European Patent. EP 1325754 B1

[52] B. Cullen, P.W. Watt, C. Lundqvist, D. Silcock, R.J. Schmidt, D. Bogan, N.D. Light. The role of oxidised regenerated cellulose/collagen in chronic wound repair and its potential mechanism of action. The International Journal of Biochemistry & Cell Biology 34 (2002) 1544-1556

[53] V. DiTizio, F. DiCosmo, Y. Xiao. Non-adhesive elastic gelatin matrices. US Patent. US 8,628,800 B2 (2014)

[54] E.J. Kim, J.S. Choi, J.S. Kim, Y.C. Choi, Y.W. Cho. Injectable and Thermosensitive Soluble Extracellular Matrix and Methylcellulose Hydrogels for Stem Cell Delivery in Skin Wounds. Biomacromolecules 17 (2016) 4-11

[55] E.J. Yun, B. Yon, M.K. Joo, B. Jeong. Cell Therapy for Skin Wound Using Fibroblast Encapsulated Poly(ethylene glycol)-poly(l-alanine) Thermogel. Biomacromolecules 13 (2012) 1106-1111

[56] B.D. Almquist, S.A. Castleberry, J.B. Sun, A.Y. Lu, P.T. Hammond. Combination Growth Factor Therapy via Electrostatically Assembled Wound Dressings Improves Diabetic Ulcer Healing In Vivo. Advanced Healthcare Materials 4 (2015) 2090-2099

[57] P. Mostafalu, G. Kiaee, G. Giatsidis, A. Khalilpour, M. Nabavinia, M.R. Dokmeci, S. Sonkusale, D.P. Orgill, A. Tamayol, A. Khademhosseini. A Textile Dressing for Temporal and Dosage Controlled Drug Delivery. Advanced Functional Materials (2017), DOI: 10.1002/adfm.201702399

[58] Q. Li, Y. Niu, H. Diao, L. Wang, X. Chen, Y. Wang, L. Dong, C. Wang. In situ sequestration of endogenous PDGF-BB with an ECM-mimetic sponge for accelerated wound healing. Biomaterials 148 (2017) 54-68

[59] X. Zhao, H. Wu, B. Guo, R. Dong, Y. Qiu, P.X. Ma. Antibacterial anti-oxidant electroactive injectable hydrogel as self-healing wound dressing with hemostasis and adhesiveness for cutaneous wound healing. Biomaterials 122 (2017) 34-47

[60] L.H.H. Olde Damink, P.J. DijkstraM. J.A. Van Luyn, P.B. Van Wachem, P. Nieuwenhuis, J. Feijen. Glutaraldehyde as a crosslinking agent for collagen-based biomaterials. Journal of Materials Science: Materials in Medicine 1995 (69) 460-472

[61] M.J. Haugh, C.M. Murphy, R.C. McKiernan, C. Altenbuchner, F.J. O'Brien. Tissue Engineering Part A 17 (2011) 1201-1208

[62] J. Qiu, J. Li, G. Wang, L. Zheng, N. Ren, H. Liu, W. Tang, H. Jiang, Y. Wang. In vitro Investigation on the Biodegradability and Biocompatibility of Genipin Cross-linked Porcine Acellular Dermal Matrix with Intrinsic Fluorescence. ACS Applied Materials and Interfaces 5 (2013) 344-350

[63] L.H.H. Olde Damink, P.J. Dijkstra, M.J. van Luyn, P.B. van Wachem, P. Nieuwenhuis, J. Feijen. Cross-linking of dermal sheep collagen using a water-soluble carbodiimide. Biomaterials 17 (1996) 765-773

[64] K. Nam, Y. Sakai, Y. Hashimoto, T. Kimura, A. Kishida. Fabrication of a heterostructural fibrillated collagen matrix for the regeneration of soft tissue function. Soft Matter 8 (2012) 472-480

[65] C. Helary, A. Abed, G. Mosser, L. Louedec, D. Letourneur, T. Coradin, M.M Giraud-Guille, A. Meddahi-Pellé. Evaluation of dense collagen matrices as medicated wound dressing for the treatment of cutaneous chronic wounds. Biomaterials Science 3 (2015) 373-382

[66] V. Natarajan, N. Krithica, B. Madhan, P.K. Sehgal. Preparation and properties of tannic acid cross-linked collagen scaffold and its application in wound healing. Journal of Biomedical Materials Research Part B: Applied Biomaterials 101B (2013) 560-567

[67] N. Ninan, A. Forget, V.P. Shastri, N.H. Voelcker, A. Blencowe. Antibacterial and Anti-Inflammatory pH-Responsive Tannic Acid-Carboxylated Agarose Composite Hydrogels for Wound Healing. ACS Applied Materials and Interfaces 8 (2016) 28511-28521

[68] A. Francesko, D. Soares da Costa, R.L. Reis, I. Pashkuleva, T. Tzanov. Functional biopolymer-based matrices for modulation of chronic wound enzyme activities. Acta Biomaterialia 9 (2013) 5216-5225

[69] T. Muthukumar, R. Senthil, T.P. Sastry. Synthesis and characterization of biosheet impregnated with *Macrotyloma uniflorum* extract for burn/wound dressings. Colloids and Surfaces B: Biointerfaces 102 (2013) 694-699

[70] D. Govindarajan, N. Duraipandy, K.V. Srivatsan, R. Lakra, P.S. Korapatti, R. Jayavel, M.S. Kiran. Fabrication of Hybrid Collagen Aerogels Reinforced with Wheat Grass Bioactives as Instructive Scaffolds for Collagen Turnover and Angiogenesis for Wound Healing Applications. ACS Applied Materials and Interfaces 9 (2017) 16939-16950

[71] T. Muthukumar, K. Anbarasu, D. Prakash, T.P. Sastry. Effect of growth factors and pro-inflammatory cytokines by the collagen biocomposite dressing material containing *Macrotyloma uniflorum* plant extract—In vivo wound healing. Colloids and Surfaces B: Biointerfaces 121 (2014) 178-188

[72] A. Yu, H. Niiyama, S. Kondo, A. Yamamoto, R. Suzuki, Y. Kuroyanagi. Wound dressing composed of hyaluronic acid and collagen containing EGF or bFGF: comparative culture study. Journal of Biomaterials Science, Polymer Edition 24 (2013) 1015-1026

[73] S. Kondo, Y. Kuroyanagi. Development of a Wound Dressing Composed of Hyaluronic Acid and Collagen Sponge with Epidermal Growth Factor. Journal of Biomaterials Science 23 (2012) 629-643.

[74] S. Kondo, H. Niiyama, A. Yu, Y. Kuroyanagi. Evaluation of aWound Dressing Composed of Hyaluronic Acid and Collagen Sponge Containing Epidermal Growth Factor in Diabetic Mice. Journal of Biomaterials Science 23 (2012) 1729-1740

[75] G. Tronci, S.J. Russell, D.J. Wood. Photo-active collagen systems with controlled triple helix architecture. Journal of Materials Chemistry B 1 (2013) 3705-3715

[76] G. Tronci, C.A. Grant, N.H. Thomson, S.J. Russell, D.J. Wood. Influence of 4-vinylbenzylation on the rheological and swelling properties of photo-activated collagen hydrogels. MRS Advances 1 (2016) 533-538

[77] G. Tronci, C.A. Grant, N.H. Thomson, S.J. Russell, D.J. Wood. Multi-scale mechanical characterization of highly swollen photo-activated collagen hydrogels. Journal of the Royal Society Interface 12 (2015) 20141079

[78] G. Tronci, J. Yin, R.A. Holmes, H. Liang, S.J. Russell, D.J. Wood. Protease-sensitive atelocollagen hydrogels promote healing in a diabetic wound model. Journal of Materials Chemistry B 4 (2016) 7249-7258

[79] N. Mao, S.J. Russell, B. Pourdeyhimi. Characterisation, testing and modelling of nonwoven fabrics (Chapter 9). Handbook of Nonwovens, first ed., Woodhead Publishing (2006).

[80] M.T. Arafat, G. Tronci, J. Yin, D.J. Wood, S.J. Russell. Biomimetic wet-stable fibres via wet spinning and diacid-based crosslinking of collagen triple helices. Polymer 77 (2015) 102-112

[81] Y.P. Kate, D.L. Christiansen, R.A. Hahn, S.-J. Shieh, J.D. Goldstein, F.H. Silver. Mechanical properties of collagen fibres: a comparison of reconstituted and rat tail tendon fibres. Biomaterials 10 (1989) 38-42

[82] M.-C. Wang, G.D. Pins, F.H. Silver. Collagen fibres with improved strength for the repair of soft tissue injuries. Biomaterials 15 (1994) 507-512

[83] D.I. Zeugolis, R.G. Paul, G. Attenburrow. Post-self-assembly experimentation on extruded collagen fibres for tissue engineering applications. Acta Biomaterialia 4 (2008) 1646-1656

[84] J.M. Caves, V.A. Kumar, J. Wen, W. Cui, A. Martinez, R. Apkarian, J.E. Coats, K. Berland, E.L. Chaikof. Fibrillogenesis in Continuously Spun Synthetic Collagen Fiber. Journal of Biomedical Materials Research Part B: Applied Biomaterials 93B (2010) 24-38

[85] X. Wang, T. Wu, W. Wang, C. Huang, X. Jin. Regenerated collagen fibers with grooved surface texture: Physicochemical characterization and cytocompatibility. Materials Science and Engineering C 58 (2016) 750-756

[86] M. Meyer, H. Baltzer, K. Schwikal. Collagen fibres by thermoplastic and wet spinning. Materials Science and Engineering C 30 (2010) 1266-1271

[87] G. Tronci, S.J. Russell. D.J. Wood. Improvements in and relating to collagen based materials. US Patent Application #14778,656 (2016)

[88] C. Haynl, E. Hofmann, K. Pawar, S. Förster, T. Scheibel. Microfluidics-Produced Collagen Fibers Show Extraordinary Mechanical Properties. Nano Letters 16 (2016) 5917-5922

[89] S. Agarwal, J.H. Wendorff, A. Greiner. Progress in the Field of Electrospinning for Tissue Engineering Applications. Advanced Materials 21 (2009) 3343-3351

[90] J. Bürck, S. Heissler, U. Geckle, M.F. Ardakani, R. Schneider, A.S. Ulrich, M. Kazanci. Resemblance of Electrospun Collagen Nanofibers to Their Native Structure. Langmuir 29 (2013) 1562-1572

[91] D.I. Zeugolis, S.T. Khew, E.S.Y. Yew, A.K. Ekaputra, YW. Tong, L.-Y.L. Yung, D.W. Hutmacher, C. Sheppard, M. Raghunath. Electro-spinning of pure collagen nano-fibres – Just an expensive way to make gelatin? Biomaterials 29 (2008) 2293-2305

[92] S. Tort, F. Acartürk, A. Beşikci. Evaluation of three-layered doxycycline-collagen loaded nanofiber wound dressing. International Journal of Pharmaceutics 529 (2017) 642-653

[93] D.W. Chen, Y.-He. Hsu, J.-Y. Liao, S.-J. Liu, J.-K. Chen, S.W.-N. Ueng. Sustainable release of vancomycin, gentamicin and lidocaine from novel electrospun sandwich-structured PLGA/collagen nanofibrous membranes. International Journal of Pharmaceutics 430 (2012) 335-341

[94] S. Singaravelu, G. Ramanathan, T. Muthukumar, M.D. Raja, N. Nagiah, S. Thyagarajan, A. Aravinthan, P. Gunasekaran, T.S. Natarajan, G.V.N.G. Selva, J.-H. Kim, U.T. Sivagnanam. Durable keratin-based bilayered electrospun mats for wound closure. Journal of Materials Chemistry B 4 (2016) 3982-3997

[95] J.H. Ko, H.Y. Yin, J. An, D.J. Chung, J.H. Kim, S.B. Lee, D.G. Pyun. Characterization of Cross-linked Gelatin Nanofibres through Electrospinning. Macromolecular Research 18 (2010) 137-143

[96] Y.Z. Zhang, J. Venugopal, Z.M. Huang, C.T. Lim, S. Ramakrishna. Crosslinking of the electrospun gelatin nanofibers. Polymer 47 (2006) 2911-2917

[97] C.H. Huang, C.Y. Chi, Y.S. Chen, K.Y. Chen, P.L. Chen, C.H. Yao. Evaluation of proanthocyanidin-crosslinked electrospun gelatin nanofibres for drug delivering system, Materials Science and Engineering C 32 (2012) 2476-2483

[98] Q. Li, Y. Niu, H. Diao, L. Wang, X. Chen, Y. Wang, L. Dong, C. Wang. In situ sequestration of endogenous PDGF-BB with an ECM-mimetic sponge for accelerated wound healing. Biomaterials 148 (2017) 54-68

[99] C. Dhand, V. A. Barathi, S.T. Ong, M. Venkatesh, S. Harini, N. Dwivedi, E.T.L. Goh, M. Nandhakumar, J.R. Venugopal, S.M. Diaz, M.H.U.T. Fazil, X.J. Loh, L.S. Ping, R.W. Beuerman, N.K. Verma, S. Ramakrishna, R. Lakshminarayanan. Latent Oxidative Polymerization of Catecholamines as

Potential Cross-linkers for Biocompatible and Multifunctional Biopolymer Scaffolds. ACS Applied Materials Interfaces 8 (2016) 32266-32281

[100] M. Zhang, H. Lin, Y. Wang, G. Yang, H. Zhao, D. Sun. Fabrication and durable antibacterial properties of 3D porous wetelectrospun RCSC/PCL nanofibrous scaffold with silver nanoparticles. Applied Surface Science 414 (2017) 52-62

[101] J.-P. Chen, W.-L. Lee. Collagen-grafted temperature-responsive nonwoven fabric for wound dressing. Applied Surface Science 255 (2008) 412-415

[102] C.-C. Wang, W.-Y. Wu, C.-C. Chen. Antibacterial and swelling properties of N-isopropyl acrylamide grafted and collagen/chitosan-immobilized polypropylene nonwoven fabrics. Journal of Biomedical Materials Research Part B: Applied Biomaterials 96B (2011) 16-24